\documentclass[useAMS,usegraphicx,usenatbib]{mn2e}
\usepackage{units}
\usepackage{amssymb}
\usepackage{amsmath}
\usepackage{hyperref}
\def\aj{AJ}%
\def\araa{ARA\&A}%
\def\apj{ApJ}%
\def\apjl{ApJ}%
\def\apjs{ApJS}%
\def\aap{A\&A}%
\def\mnras{MNRAS}%
\def\nar{New A Rev.}%
\def\pasj{PASJ}%
\def\nat{Nature}%
\def\physrep{Phys.~Rep.}%
\title[Cosmic reionization of hydrogen and helium]{Cosmic reionization of hydrogen and helium: contribution
from both mini-quasars and stars}

\author[J.-M. Hao, Y.-F. Yuan and L. Wang]{Jing-Meng Hao$^{1,2}$\thanks{E-mail:
haojm@mail.ustc.edu.cn; yfyuan@ustc.edu.cn; leiwang@pmo.ac.cn}, Ye-Fei Yuan$^{2}$\footnotemark[1] and Lei Wang$^{3,4}$\footnotemark[1] \\
$^{1}$Center for Astrophysics, Guangzhou University, Guangzhou 510006, China\\
$^{2}$Key Laboratory for Research in Galaxies and Cosmology CAS, Department of Astronomy,\\
University of Science and Technology of China, Hefei, Anhui 230026, China\\
$^{3}$Purple Mountain Observatory, the Partner Group of MPI f\"{u}r
Astronomie, 2 West Beijing Road, Nanjing 210008, China\\
$^{4}$Key Laboratory for Research in Galaxies and Cosmology,\\
Shanghai Astronomical Observatory, Nandan Road 80, Shanghai 200030, China}

\begin{document}

\pagerange{\pageref{firstpage}--\pageref{lastpage}} \pubyear{2015}

\maketitle

\label{firstpage}

\begin{abstract}
Observations on the high-redshift galaxies at $z>6$ imply that their ionizing
emissivity is unable to fully reionize the Universe at $z\sim 6$. Either a high escape fraction of ionizing photons
from these galaxies or a large population
of faint galaxies below the detection limit are required. However, these requirements are somewhat in tension
with present observations. In this work, we explored the combined contribution
of mini-quasars and stars to the reionization of cosmic hydrogen and helium.
Our model is roughly consistent with:
(1) the low escape fractions of ionizing photons from the observed galaxies,
(2) the optical depth of Cosmic Microwave Background (CMB) measured
by the WMAP-7, and
(3) the redshift of the end of hydrogen and helium reionization
at $z\approx 6$ and $z\approx 3$, respectively.
Neither an extremely high escape fraction nor a large population of fainter galaxies
is required in this scenario.
In our most optimistic model, more than $\sim20\%$ of the cosmic helium is reionized by $z\sim6$,
and the ionized fraction of cosmic helium rapidly climbs to more than $50\%$ by $z\sim5$.
These results may imply that better measurements of helium reionization, especially at high redshifts, could be helpful in
constraining the growth of intermediate-mass black holes (IMBHs) in the early Universe,
which would shed some light on the puzzles concerning the formation of supermassive black holes (SMBHs).
\end{abstract}


\section{Introduction}

One major focus in the present cosmology is the reionization of the intergalactic
medium (IGM) in the Universe. Observations of the Cosmic Microwave
Background (CMB) from the 7-yr \emph{Wilkinson Microwave Anisotropy
Probe} (WMAP-7) measure the Thomson electron scattering optical depth
$\tau=0.088\pm0.015$, suggesting that hydrogen reionization begins
no later than $z=10.6\pm1.2$ \citep{2011ApJS..192...18K}.
Additional pieces of evidence come from the absorption
spectra of high-redshift quasars \citep{2006ARA&A..44..415F,2011Natur.474..616M,2011MNRAS.416L..70B,2013MNRAS.428.3058S}
and the rapid evolution of the luminosity function of Ly$\alpha$ emitters at $z>6$ \citep{2010ApJ...723..869O,2012ApJ...744..179S},
implying an end to hydrogen reionization at $6\lesssim z\lesssim7$.
All these constraints suggest that hydrogen reionization is likely
to be a much more extended process in the redshift range $6<z<12$
\citep[e.g.][]{2003ApJ...586..693W,2006MNRAS.371L..55C,2007MNRAS.376..534I,2012ApJ...746..125H,2013ApJ...768...71R}.

Among various possible ionizing sources, stars in galaxies are commonly thought to be the most
likely candidates \citep[e.g.][]{2008ApJ...682L...9F,2010Natur.468...49R}.
However, the ionizing emissivity of the observed galaxies at $z>6$
seems insufficient to maintain the cosmic hydrogen in a fully ionized state
at $z\sim6$ unless the escape fraction of ionizing photons from
these galaxies is larger than $\sim30\%$, otherwise a large population
of faint galaxies below the detection limit would be required
\citep{2007MNRAS.382..325B,2010MNRAS.409..855B,2010ApJ...714L.202T,2012ApJ...759L..38A,2012ApJ...752L...5B,2012MNRAS.425.1413F,2012ApJ...758...93F,2012MNRAS.423..862K,2013ApJ...768...71R,2014MNRAS.442.2560W}.
The requirement for such a high escape fraction at $z\gtrsim6$ is somewhat in tension with current observational
constraints on the escape fraction at lower redshifts
\citep{2001ApJ...558...56H,2009ApJ...692.1287I,2010ApJ...725.1011V,2011ApJ...736...18N,2012ApJ...751...70V}.
Although there exists some evidence supporting the redshift evolution
of the escape fraction \citep{2006ApJ...651L..89R,2011A&A...530A..87P,2013MNRAS.431.2826F,2013MNRAS.428L...1M,2014MNRAS.438.2097F},
which is still inconclusive.
On the other hand, some theoretical studies suggest that the star formation activity in
faint galaxies might be strongly suppressed \citep{2012ApJ...753...16K,2012ApJ...749...36K}.

Introducing the contribution from other ionizing sources might
partially solve this so-called ``photon-starved'' problem. Although it is
widely believed that the contribution of quasars to the ionizing radiation
decreases rapidly from $z\sim3$ to $z\sim6$ and
becomes negligible at $z\gtrsim 6$ \citep[e.g.][]{2001AJ....122.2833F,2009ApJ...692.1476C,2010AJ....139..906W},
the contribution from mini-quasars at high redshifts could still play an important role. The existence
of black holes (BHs) as massive as $\sim10^{9}\,\mathrm{M_{\odot}}$ at
$z\sim7$ becomes reasonable only when starting from a seed BH
\footnote{The origin of the initial seed BHs remains an open question. The most obvious candidate is the collapse
of massive Population III stars at $z\gtrsim 20$ \citep[e.g.][]{2001ApJ...551L..27M,2001ApJ...552..459H,2009ApJ...696.1798T},
leaving behind remnant BHs with masses $\gtrsim 100\,\mathrm{M_{\odot}}$.
An alternative possibility is the direct collapse of primordial gas into a $10^4-10^6\,\mathrm{M_{\odot}}$ BH in halos
with $T_{\mathrm{vir}}\gtrsim 10^4\,\mathrm{K}$ \citep[e.g.][]{2002ApJ...569..558O,2003ApJ...596...34B,2006MNRAS.371.1813L,2014MNRAS.439.1092T}.},
which increases its mass via gas accretion or/and merging with each other. This scenario implies that there should be plenty of
intermediate-mass black holes (IMBHs) with masses of the order of $10^{2}-10^{5}\,\mathrm{M_{\odot}}$ shining at $z\gtrsim 7$.
The recent identification of IMBHs also provides strong evidence for
this scenario \citep{2009Natur.460...73F,2013A&A...552A..49L}. If
such IMBHs accrete as mini-quasars at high redshifts, they could be
important sources of the early reionization of the IGM. A number of authors
\citep[e.g.][]{2004ApJ...604..484M,2004ApJ...613..646D,2007MNRAS.375.1269Z} argued
that mini-quasars, consistent with the constraint from the soft X-ray background (SXRB),
could still provide a relevant contribution to the reionization of
the IGM at $z\gtrsim 6$.

Recently, by exploring the high-redshift $M_{\mathrm{bh}}-\sigma_{*}$
relation in light of the numerical simulation
of the merger tree of dark matter halos and using the Press-Schechter (PS)
formalism, \citet{2010RAA....10..199W} calculated the central IMBH mass density
as a function of redshift. And furthermore, by combining it with a parameterized UV
photon emission efficiency, they found that the contribution of mini-quasars
to hydrogen reionization lies in the range $\sim25\%-50\%$ at $z\sim6$.
However, they did not consider the impact of the helium ionizing radiation of these mini-quasars
on the IGM reionization and ignored the contribution from stars.
In this paper, we adopt the same BH mass density to estimate the number of
ionizing photons from mini-quasars, assuming a power-law spectrum for the BH emission.
In combination with the fitting star formation rate density
from \citet{2008MNRAS.388.1487L}, we assess the combined contribution of
mini-quasars and stars to the reionization of cosmic hydrogen and helium.

The structure of this paper is as follows. In Sect.~2, we present
the formalism for computing the reionization histories
of hydrogen and helium and the number of ionizing photons
from different types of sources. The implications of our results are
discussed in Sect.~3, and our conclusions are summarized in Sect.~4.
The cosmological parameters used in this paper are from the WMAP-7 results \citep{2011ApJS..192...18K}: $\Omega_{\mathrm{m}}=0.266$,
$\Omega_{\mathrm{\Lambda}}=0.734$, $\Omega_{\mathrm{b}}=0.0449$,
$h=0.71$, $\sigma_{8}=0.801$. We use $n=1$ for a scale-invariant primordial power spectrum.

\section{The reionization history of hydrogen and helium}

The X-ray or UV emission from (mini-)quasars and stars can ionize the
neutral hydrogen (H\,I), helium (He\,I) or singly ionized helium
(He\,II) in the IGM. In order to estimate the evolution of the
regions of the ionized hydrogen and helium during the epoch of reionization,
we follow the straightforward considerations of \citet{2001PhR...349..125B}.
Given the number density of ionizing photons from stars and (mini-)quasars
in a homogeneous but clumpy Universe, the evolution of the volume
filling factor $Q_{i}$ (for species $i=\mathrm{HII,\, HeII,\, HeIII}$)
of the ionized regions can be described by
\begin{equation}
\frac{\mathrm{d}Q_{\mathrm{i}}}{\mathrm{d}t}=\frac{1}{n^{0}_{i}(1+z)^{3}}\frac{\mathrm{d}N_{\gamma, i}}{\mathrm{d}t}-\alpha_{\mathrm{B,i}}Cn_{e}(1+z)^{3}Q_{\mathrm{i}},
\end{equation}
where $n^{0}_{i}$ is the comoving density, and
$n_{\mathrm{H}}^{0}=\unit[1.88\times10^{-7}(\Omega_{\mathrm{b}}h^{2}/0.022)]{cm^{-3}}$
and $n_{\mathrm{He}}^{0}=\unit[0.19\times10^{-7}(\Omega_{\mathrm{b}}h^{2}/0.022)]{cm^{-3}}$,
respectively. Here $n_{e}$ is the number density of electrons,
$\alpha_{\mathrm{B}}$ is the case B recombination coefficient
and $C\equiv\left\langle n_{\mathrm{H}}^{2}\right\rangle /\bar{n}_{\mathrm{H}}^{2}$
is the volume-averaged clumping factor of the IGM.
The recombination coefficients for H\,II, He\,II and He\,III at $T\backsimeq10^{4}\,\mathrm{K}$
are adopted as follows \citep{2003ApJ...586..693W}:
$\alpha_{\mathrm{B,HII}}=\unit[2.6\times10^{-13}]{cm^{3}\, s^{-1}}$,
$\alpha_{\mathrm{B,HeII}}=\unit[2.73\times10^{-13}]{cm^{3}\, s^{-1}}$ and
$\alpha_{\mathrm{B,HeIII}}=\unit[13\times10^{-13}]{cm^{3}\, s^{-1}}$,
while for the clumping factor $C$, we use the following simple form
given by \citet{2006ApJ...650....7H}: $C(z)=1+9(7/(1+z))^{2}$ for
$z\geq6$ and $C=10$ for $z<6$. The number densities of ionizing
photons for H\,II, He\,II and He\,III are calculated through the integration
of the corresponding spectra in the energy intervals $13.6<E<24.6\,\mathrm{eV}$,
$24.6<E<54.4\,\mathrm{eV}$ and $E>54.4\,\mathrm{eV}$, respectively.
Two classes of sources, mini-quasars and stars, are considered as
the contributors to the ionizing photons, which will be discussed in the
following sections. We also note that, following \citet{2003ApJ...586..693W},
two constraints are considered in this model, namely, (1) $Q_{\mathrm{HeIII}}\leq Q_{\mathrm{HII}}$;
and (2) $Q_{\mathrm{HeII}}\leq Q_{\mathrm{HII}}-Q_{\mathrm{HeIII}}$,
showing that the helium ionization front never overtakes the hydrogen ionization
front.

The optical depth of CMB photons due to the Thomson Scattering
is given by
\begin{equation}
\tau_{e}=\int\mathrm{d}z\frac{\mathrm{d}t}{\mathrm{d}z}c\sigma_{\mathrm{T}}n_{e}(z)(1+z)^{3},
\end{equation}
where $\sigma_{\mathrm{T}}=6.65\times10^{-25}\,\mathrm{cm^{-2}}$
is the cross section of Thomson scattering and $n_{e}$ is the
number density of electron.

\subsection{Ionizing photons from mini-quasars}

In this section, we derive the number of ionizing photons from the central accreting IMBHs, shining as mini-quasars.
The ionizing radiation from accreting BHs primarily depends on the following
three quantities: the redshift evolution of the mass density of BHs,
the emission spectrum of the accreting BH and the duty cycle of BHs.

Our calculation for the mass density of central BHs as a function of redshift closely follows
the simple model of \citet{2010RAA....10..199W} which is based on the PS mass function
of dark matter halos. Here we only briefly describe the key features of this model, and refer the reader to that paper for more details.
We assume that, at the center of each dark matter halo,
there exists a BH with a mass ($M_{\mathrm{BH}}$) that is correlated with the host halo mass.
Then the mass density of BHs could be estimated
simply by calculating the number density of halos
as a function of redshift. Here we adopt
the modified version of PS mass function proposed by \citet{1999MNRAS.308..119S}
to estimate the number of dark matter halos, which is
\begin{eqnarray}
n_{\mathrm{ST}}(M,z)\,\mathrm{d}M & = & A\sqrt{\frac{2a_{1}}{\pi}}\frac{\rho_{\mathrm{m}}}{M}\left[1+\left(\frac{\sigma^{2}(M,z)}{a_{1}\delta_{c}^{2}}\right)^{p}\right]\frac{\delta_{c}}{\sigma(M,z)}\nonumber \\
 &  & \exp\left[-\frac{a_{1}\delta_{c}^{2}}{2\sigma^{2}(M,z)}\right]\frac{\mathrm{d}\ln\sigma^{-1}(M,z)}{\mathrm{d}M}\,\mathrm{d}M,
\end{eqnarray}
where $A=0.3222$, $a_{1}=0.707$, $p=0.3$, $\delta_{c}=1.686$ and
$\rho_{\mathrm{m}}$ is the current mean density of the Universe.
The deviation of the linear density field $\sigma(M,z)$ is given by
\begin{equation}
\sigma^{2}(M,z)=\frac{D^{2}(z)}{2\pi^{2}}\int_{0}^{\infty}k^{2}P(k)W^{2}(k,M)\,\mathrm{d}k,
\end{equation}
where $W(k,M)$ is the top-hat filter, and $D(z)$ is the growth factor of linear perturbations.

The mass of the central BH ($M_{\mathrm{BH}}$) is assumed to be correlated with the the stellar
velocity dispersion $\sigma_{*}$ through the relation
\begin{equation}
\log M_{\mathrm{BH}}=\theta+\phi\log(\sigma_{*}/\sigma_{0}) \,\,\, ,
\end{equation}
where $\sigma_{0}=200\,\mathrm{km\, s^{-1}}$.
Basing on the improved self-adaptive merger tree model of dark matter halos \citep{2008AIPC..968..369Y},
\citet{2010RAA....10..199W} explored the $M_{\mathrm{BH}}-\sigma_{*}$ relation at high redshifts
with the following BH formation scenario: (i) each halo is randomly assigned a seed BH; (ii) seed BHs as remnants of Pop III stars
have a slightly top-heavy initial mass function in the mass range $200\mathrm{M_{\odot}}<M_{\mathrm{BH}}<1000\mathrm{M_{\odot}}$
\footnote{Although recent studies imply a smaller characteristic mass of Pop III stars,
it is yet inconclusive how this scenario would affect the growth of MBHs.
Besides, several authors \citep[e.g.][]{2003ApJ...582..559V} also argued that the evolution of MBHs is not very sensitive to the initial mass of seed BHs.
Thus, we prefer a traditional seed BH model in this work.};
(iii) the mass growth of BHs is considered to be dependent on gas accretion during halo mergers, and the mass accretion rate is
$\dot{M}_{\mathrm{BH}}=M_{\mathrm{BH}}/t_{\mathrm{ef}}$, where $t_{\mathrm{ef}}$ is the Salpeter $e$-folding time;
(iv) the free parameters, such as the radiative efficiency and Eddington ratio,
are constrained to make sure the final BH mass agrees with the local $M_{\mathrm{BH}}-\sigma_{*}$ relation.
They found that the $M_{\mathrm{BH}}-\sigma_{*}$ relation with parameters $\theta=6.67$
and $\phi=2.79$ can provide a good fit to the numerical results at $z>6$.
The stellar velocity dispersion $\sigma_{*}$ is related to the circular velocity
of halos $V_{\mathrm{c}}$ by \citep{2002ApJ...578...90F,2005ApJ...631..785P}
\begin{equation}
\log V_{\mathrm{c}}=0.84\log\sigma_{*}+0.55.
\end{equation}
The relation between the circular velocity $V_{\mathrm{c}}$ and the
halo mass $M$ is given by \citep{2001PhR...349..125B}
\begin{equation}
V_{\mathrm{c}}=23.4\left(\frac{M}{10^{8}h^{-1}\mathrm{M_{\odot}}}\right)^{1/3}\left[\frac{\Omega_{\mathrm{m}}}{\Omega_{\mathrm{m}}^{z}}\frac{\Delta_{\mathrm{c}}}{18\pi^{2}}\right]^{1/6}\left(\frac{1+z}{10}\right)^{1/2}\mathrm{km\, s^{-1}}
\end{equation}
with
\begin{equation}
\Delta_{\mathrm{c}}=18\pi^{2}+82d-39d^{2}
\end{equation}
and $d\equiv\Omega_{\mathrm{m}}(z)-1$.

\begin{figure*}
\includegraphics[width=84mm]{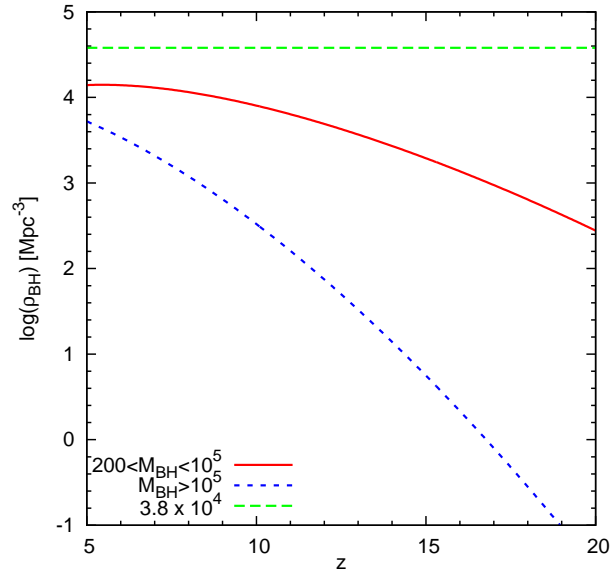}
\caption{Evolution of the mass density of central BHs as a function of redshift.
The solid line and short-dashed line represent the mass density of
IMBHs and MBHs, respectively. The dashed line is the constraints on the BH
mass density from the SXRB derived by \citet{2005MNRAS.362L..50S}.}
\label{fig1}
\end{figure*}

In Fig.~\ref{fig1},
we show the BH mass densities as a function of redshift for IMBHs and massive BHs (MBHs; $>10^{5}\,\mathrm{M_{\odot}}$),
respectively.
We henceforth refer to accreting MBHs as ``quasars''.
For comparison, we also plot
the strong upper limit on the IMBH density at
$z>6$ of $\rho_{\mathrm{IMBH}}<3.8\times10^{4}\,\mathrm{M_{\odot}\, Mpc^{-3}}$ derived by \citet{2005MNRAS.362L..50S},
which is based on the measurement of the unresolved fraction of the SXRB.
As can be seen from Fig.~\ref{fig1}, the result is well below the
constraints from the SXRB.

Although the exact emission spectrum of mini-quasars remains poorly understood,
its typical spectrum is expected to be harder than that of quasars.
As suggested by observations of ultraluminous X-ray sources (possibly associated to IMBHs; \citealt{2003ApJ...585L..37M}) in nearby galaxies,
the mini-quasar spectrum could be approximatively described by two continuum components,
including both a ``multicolor disk blackbody'' \citep{1984PASJ...36..741M} component and
a non-thermal power-law component $L_{\mathrm{E}}\propto E^{-\alpha}$ with $\alpha\approx1$.
To simplify the calculation, we follow previous studies \citep{2005MNRAS.363.1069K,2006NewAR..50..204D,2007MNRAS.375.1269Z}
and assume that the mini-quasar spectrum has a simple power-law form.
Specifically, two types of power-law spectral energy distributions (SEDs) are considered: the High Energy (HE) case with the energy range
$200\,\mathrm{eV}<E<100\,\mathrm{keV}$ and the Low Energy (LE) case with a low-energy cutoff of $10.4\,\mathrm{eV}$.
Note that, the HE case assumes that all the ionizing UV photons are absorbed in the close vicinity of the source.
As a comparison, we also note that the typical spectral
index of quasars is measured to be $\alpha=1.7$
\citep[e.g.][]{2002ApJ...565..773T,2006A&A...451..457T}.

Then the number of hydrogen (helium) ionizing photons from accreting IMBHs can be evaluated by
\begin{equation}
\frac{\mathrm{d}N_{\gamma}}{\mathrm{d}t}=\frac{\int_{E_{\mathrm{min}}}^{E_{\mathrm{max}}}F(E)\frac{\mathrm{d}E}{E}}{\int F(E) \mathrm{d}E}\epsilon\bar{L}P,
\end{equation}
where $\epsilon$ is the Eddington ratio, $P$ represents the duty cycle and
$\bar{L}$ is the total luminosity emitted by accreting IMBHs per
unit of comoving volume:
\begin{eqnarray}
\bar{L} & = & \int_{200}^{10^{5}}N(M(M_{\mathrm{BH}},z),z)L_{\mathrm{Edd}}(M_{\mathrm{BH}})\mathrm{d}M_{\mathrm{BH}}\nonumber \\
 & = & 1.26\times10^{38}\frac{\rho_{\mathrm{IMBH}}(z)}{\mathrm{M_{\odot}}}\,\mathrm{erg\, s^{-1}\, Mpc^{-3}},
\end{eqnarray}
where $L_{\mathrm{Edd}}$ is the Eddington luminosity, $L_{\mathrm{Edd}}(M_{\mathrm{BH}})=1.26\times10^{38}(M_{\mathrm{BH}}/\mathrm{M_{\odot}})\,\mathrm{erg\, s^{-1}}$,
and $P$ is the duty cycle of BHs, i.~e. the probability that a BH is in the active state.
A number of studies suggest that the duty cycle increases rapidly with increasing redshift \citep[e.g.][]{2006ApJ...647L..17W,2007AJ....133.2222S,2014CQGra..31x4005T}.
From the measurements of clustering of quasars, \citet{2010ApJ...718..231S} suggested that
$P\sim0.2$, 0.5 and 0.9 at $z=3.1$, 4.5 and 6, respectively.
To match these results, we therefore adopt a parameterized form of $P(z)=0.03+0.97(z/z_{\mathrm{tran}})^{\beta}/[1+(z/z_{\mathrm{tran}})^{\beta}]$ with $\beta=2$ and
$z_{\mathrm{tran}}=4$. To be optimistic, we also assume that all the ionizing photons produced by accreting BHs
escape the host halo, i.~e. $f_{\mathrm{esc,BH}}=1$.
Based on the above assumptions and descriptions, consequently,
it is straightforward to calculate the number of ionizing photons emitted
at a particular energy range $[E_{\mathrm{min}},\, E_{\mathrm{max}}]$
from accreting IMBHs per unit time.

\subsection{Ionizing photons from stars}

In order to determine the number of ionizing photons produced by stars,
the high-redshift star formation rate (SFR) density is estimated
according to \citet{2008MNRAS.388.1487L}, in which
\citet{2006ApJ...651..142H}'s compilation on the SFR density
was updated with the observations by
\citet{2004ApJ...600L.103G}, \citet{2004MNRAS.355..374B},
\citet{2004ApJ...611..660O}, \citet{2008ApJS..175...48R} and \citet{2008ApJ...686..230B}.
Basing on these observational data, \citet{2008MNRAS.388.1487L} obtained
a piecewise-linear fitting, as in \citet{2006ApJ...651..142H}, which is
\begin{equation}
\log\dot{\rho}_{*}(z)=a+b\log(1+z),
\end{equation}
where
\begin{equation}
(a,b)=\begin{cases}
(-1.70,3.30), & z<0.993\\
(-0.727,0.0549), & 0.993<z<3.80\\
(2.35,-4.46), & z>3.80
\end{cases},
\end{equation}
and $\dot{\rho}_{*}$ is the SFR in units of $\mathrm{M_{\odot}\, yr^{-1}\, Mpc^{-3}}$.
Then the number of ionizing photons is calculated by
\begin{equation}
\frac{\mathrm{d}N_{\gamma}}{\mathrm{d}t}=n_{\gamma}f_{\mathrm{esc}}\dot{\rho}_{*}(z),
\end{equation}
where $f_{\mathrm{esc}}$ represents the escape fraction of ionizing
photons from stars and $n_{\gamma}$ gives the number
of photons emitted per unit mass of stars.
Unfortunately, the escape fraction of stellar ionizing photons
from high-redshift galaxies remains poorly constrained, making this parameter one
of the primary goals of many theoretical and observational studies.
A number of studies \citep[e.g.][]{2009ApJ...692.1287I,2010ApJ...725.1011V,2011ApJ...736...18N}
have measured the escaping ionizing photons at $z\sim3-4$, suggesting that
the escape fraction is between 0.05 and 0.2.
Although the samples are small, almost all the average escape fractions
of ionizing photons from the observed massive galaxies
are less than 0.2.
By integrating over appropriate energy ranges, the parameter $n_{\gamma}$ gives
$(8.05,2.62,0.01)\times10^{60}\,\mathrm{M_{\odot}^{-1}}$ for H\,II,
He\,II, He\,III, respectively, as suggested by \citet{2005MNRAS.361..577C}.

\begin{figure*}
\includegraphics[width=84mm]{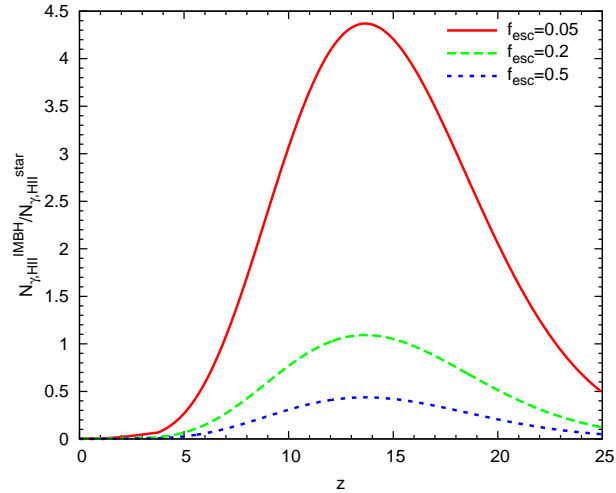}
\caption{Redshift evolution of the ratio between the IMBH and star contributions to the hydrogen ionizing photons,
assuming the LE SED ($10.4\,\mathrm{eV}<E<100\,\mathrm{keV}$) for the IMBH emission.
The escape fractions of ionizing photons from stars are taken to be $f_{\mathrm{esc}}=0.05$, $0.2$ and $0.5$ for the solid, dashed and short-dashed lines, respectively.}
\label{fig2}
\end{figure*}

For comparison,
we show how the ratio $\dot{N}_{\gamma,\mathrm{HII}}^{\mathrm{IMBH}}/\dot{N}_{\gamma,\mathrm{HII}}^{\mathrm{star}}$
evolves as a function of redshift for three different $f_{\mathrm{esc}}$ in Fig.~\ref{fig2}.
Here we consider only the LE case for the IMBH spectrum, as for the HE case, the hydrogen ionizing photons from mini-quasars are simply zero.
It is also worth stressing that all the ionizing photons from mini-quasars are assumed to be available for ionizing the IGM.
As shown in Fig.~\ref{fig2}, the ratio peaks at $z\sim13$ for all three models, decreases rapidly at lower redshifts and becomes negligible at $z\lesssim 5$.
Only for the model with $f_{\mathrm{esc}}<0.2$,
mini-quasars have a higher ionizing flux contribution than that of stars at $z\gtrsim 6$.

\section{Results}

\begin{figure*}
\includegraphics[width=84mm]{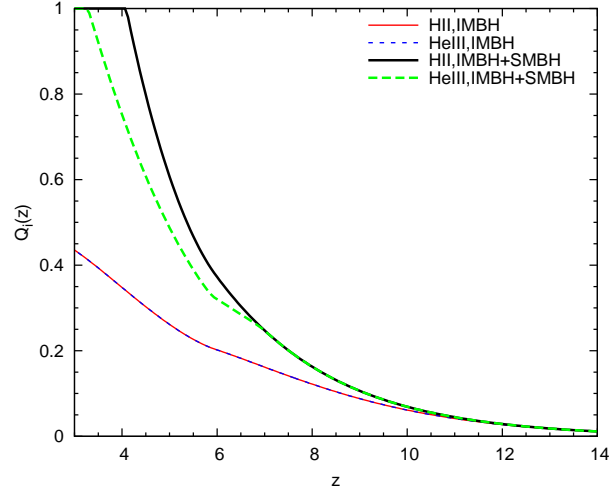}
\caption{Redshift evolution of the volume filling factor of $Q_{\mathrm{HII}}$
and $Q_{\mathrm{HeIII}}$ due to the ionizing photons from
IMBHs and MBHs, assuming the LE SED ($10.4\,\mathrm{eV}<E<100\,\mathrm{keV}$) for the IMBH emission.}
\label{fig3}
\end{figure*}

We first consider the contribution of mini-quasars, i.~e.
accreting IMBHs alone to the reionization of cosmic hydrogen and helium.
Fig.~\ref{fig3} shows the evolution histories of the filling factor of the ionized regions $Q_{\mathrm{HII}}$
(solid line) and $Q_{\mathrm{HeIII}}$ (dashed line) due to the accreting IMBHs alone.
Note that only the LE case for the IMBH emission is shown here,
since accreting IMBHs only provide a negligible contribution to hydrogen reionization for the HE case as mentioned above.
As can be seen, the accreting IMBHs can only contribute
$\sim20\%$ of hydrogen ionizing photons at $z\sim6$. The difference
between the evolution of $Q_{\mathrm{HeIII}}$ and that of $Q_{\mathrm{HII}}$
is very subtle, because of the assumption that the He~III ionization
front never overtakes the hydrogen ionization front, i.\,e. $Q_{\mathrm{HII}}\geq Q_{\mathrm{HeIII}}$.
This result implies that, in an optimistic case (LE case), the contribution of mini-quasars to hydrogen reionization
is not negligible at $z\sim6$, but still insufficient to provide
the required number of ionizing photons, consistent with previous
studies \citep{2004ApJ...604..484M,2004ApJ...613..646D,2010RAA....10..199W,2012MNRAS.426.1349M}.

As a reference, the contribution from accreting MBHs (``quasars'') is also considered.
Note that we set the emission spectrum of quasars with a spectral
index of $\alpha=1.7$ and ionizing UV and X-ray photons (i.~e. energy range of $10.4\,\mathrm{eV}<E<100\,\mathrm{keV}$).
As shown in Fig.~\ref{fig3},
only at relatively low redshifts ($z\lesssim 6$) does the contribution from MBHs play a relevant role in reionizing the IGM.
When considering the contributions from both IMBHs and MBHs,
we find that helium reionization ends at $z\approx3$,
consistent with recent observations of helium reionization
\citep[e.g.][]{2004ApJ...605..631Z,2009ApJ...704L..89M,2010ApJ...714..355F,2010ApJ...722.1312S,2011MNRAS.410.1096B,2011ApJ...733L..24W,2012MNRAS.423....7M}.
Moreover, in this scenario, the middle of helium reionization occurs at $z\sim5$ and hydrogen reionization ends at $z\approx4$.
It is worth stressing that, for simplicity, we have assumed the same $M_{\mathrm{BH}}-\sigma_{*}$
relation at all redshifts. This almost certainly constitutes an underestimate for the contribution
of accreting BHs at low redshifts and delays the completion of
helium reionization.

\begin{figure*}
\includegraphics[width=84mm]{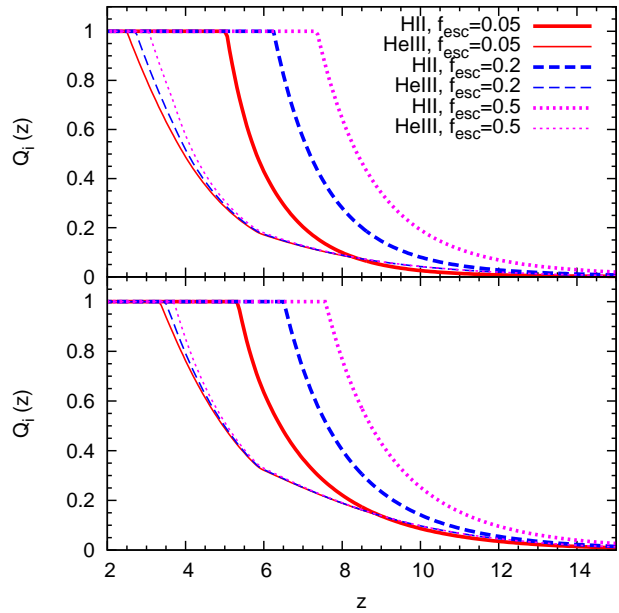}
\caption{Redshift evolution of the volume filling factor of $Q_{\mathrm{HII}}$
and $Q_{\mathrm{HeIII}}$ for three different escape fractions $f_{\mathrm{esc}}$ of ionizing photons from stars.
Here, both accreting BHs and stars are included. The IMBH spectrum has a spectral index of $\alpha=1$.
The upper panel refers to the HE case for the BH emission, while the lower panel refers to the LE case.}
\label{fig4}
\end{figure*}

We then consider the combined contribution from accreting BHs and stars
to the reionization of hydrogen and helium. Here, for the ionizing emission
from accreting BHs, both IMBHs and MBHs are included.
Given the uncertainty in the escape fraction of ionizing photons from
stars, we present the results of modeling the reionization histories
of hydrogen and helium for three different choices of the escape fraction of ionizing photons from stars
($f_{\mathrm{esc}}=0.05;\,0.2;\,0.5$). Fig.~\ref{fig4} shows
the evolution of both $Q_{\mathrm{HII}}$ and $Q_{\mathrm{HeIII}}$.
The upper panel refers to the HE case for the IMBH emission, while the lower panel is for the LE case.
It is clear that the Universe is reionized earlier for the LE case.
As seen in both panels,
the escape fraction $f_{\mathrm{esc}}$ of ionizing photons from stars does not have strong influence on the
evolution of $Q_{\mathrm{HeIII}}$, because stars
provide a negligible contribution to the helium ionizing photons.

\begin{figure*}
\includegraphics[width=84mm]{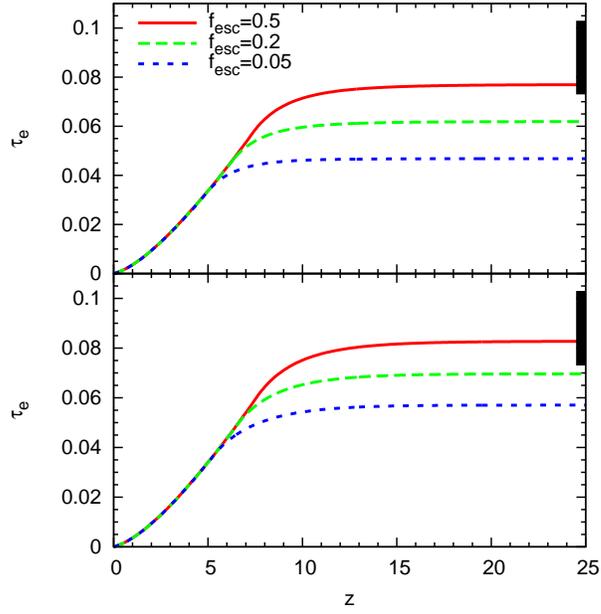}
\caption{Redshift evolution of the Thomson scattering optical depth
for three different escape fractions $f_{\mathrm{esc}}$.
The upper and lower panels refer to the HE and LE cases for the IMBH emission, respectively.
The black bar represents the WMAP-7 result, i.e., $\tau_{e}=0.088\pm0.015$.}
\label{fig5}
\end{figure*}

In the HE case, helium reionization ends at $z\approx2.7$, consistent with
observations, while in the LE case the end redshift of helium reionization $z\approx3.5$ is much earlier.
Models with $f_{\mathrm{esc}}=0.2$ and $f_{\mathrm{esc}}=0.5$ satisfy the observational constraints of the Gunn-Peterson trough (i.~e. hydrogen reionization at
$6\lesssim z_{\mathrm{reion}}\lesssim7$). However, when considering the constraints
on the CMB optical depth, only the model with $f_{\mathrm{esc}}=0.5$ is consistent with the WMAP-7 result of $\tau_{e}=0.088\pm0.015$, as shown in Fig.~\ref{fig5}.

In the LE case, only the model with $f_{\mathrm{esc}}=0.2$ is consistent with observations of the Gunn-Peterson trough, whereas hydrogen reionization
in the other two models is completed either too early ($z\gtrsim7$) or too late
($z\lesssim5.5$). However, although the model with $f_{\mathrm{esc}}=0.2$
is able to reionize the Universe at $z\sim6$, it is not able to satisfy the constraints
on the CMB optical depth simultaneously. This result imply that, to simultaneously satisfy
the joint constraints from the Gunn-Peterson test and the CMB optical depth,
an escape fraction $f_{\mathrm{esc}}$ within the range from 0.2 to 0.5 is needed in this case.

\begin{figure*}
\includegraphics[width=84mm]{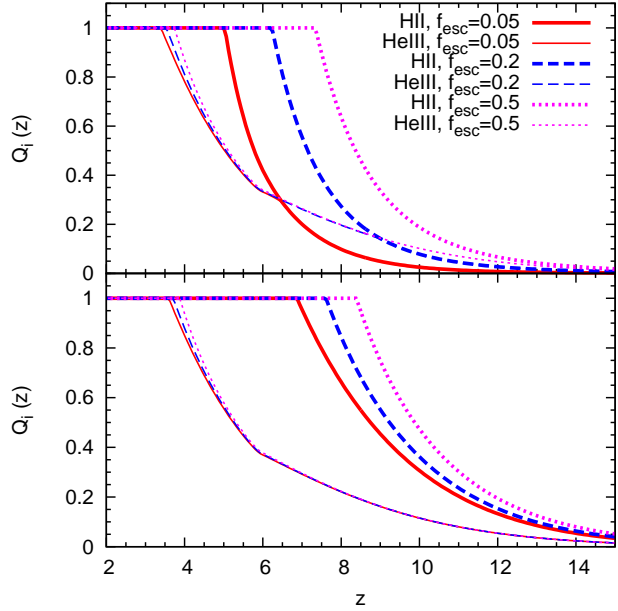}
\caption{Redshift evolution of the volume filling factor of $Q_{\mathrm{HII}}$
and $Q_{\mathrm{HeIII}}$, assuming the mini-quasar emission with a quasar-like spectrum ($\alpha=1.7$).
The upper and lower panels refer to the HE and LE cases for the IMBH emission, respectively.
Here, He\,II reionization completes much earlier at $z\gtrsim3.5$ for both HE and LE cases.}
\label{fig6}
\end{figure*}

\begin{figure*}
\includegraphics[width=84mm]{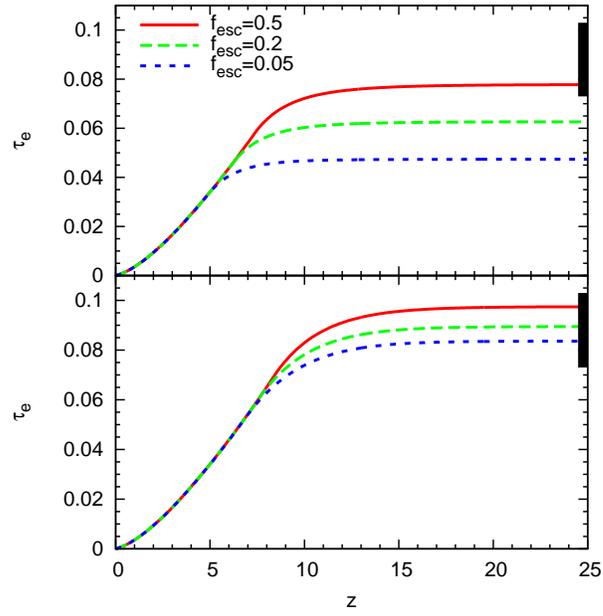}
\caption{Redshift evolution of the Thomson scattering optical depth,
assuming the mini-quasar emission with a quasar-like spectrum ($\alpha=1.7$).
The upper and lower panels refer to the HE and LE cases for the IMBH emission, respectively.
The black bar represents the WMAP-7 result, i.e., $\tau_{e}=0.088\pm0.015$.}
\label{fig7}
\end{figure*}

So far, we have analyzed the evolution of the reionization histories
for different escape fractions, assuming a power-law
spectral index of $\alpha=1$ for the ionizing emission of accreting IMBHs.
For comparison, we also explore the possibility of a quasar-like spectral index of $\alpha=1.7$
for the emission spectrum of IMBHs. The results are shown in Fig.~\ref{fig6}
and Fig.~\ref{fig7}. As seen in Fig.~\ref{fig6},
helium reionization ends much earlier at $z\gtrsim3.5$ for both HE and LE cases.
For the HE case, the results turn out to be similar to those found in the $\alpha=1$ scenario,
except that helium reionization ends much earlier at $z\gtrsim3.5$. On the other hand, for the LE case,
all three models with different $f_{\mathrm{esc}}$ can produce enough ionizing photons to obtain the CMB optical depth, as shown in Fig.~\ref{fig7}.
In this extreme case, only an escape fraction of $f_{\mathrm{esc}}\lesssim0.05$ is large enough.
These results show that, if high-redshift star-forming galaxies have an escape fraction larger than $0.05$,
then accreting IMBHs with a quasar-like spectrum would produce too many ionizing photons to be more or less in tension with observations.

In the most optimistic model, mini-quasars are able to reionize more than $\sim20\%$ of the cosmic helium by redshift $z\sim6$,
and the ionized fraction of cosmic helium rapidly climbs to more than $50\%$ by redshift $z\sim5$ when considering both IMBHs and MBHs.
These results may imply that, in addition to the constraint from the SXRB,
better measurements of the evolution history of helium reionization, especially at high redshifts,
would also be helpful in placing limits on the growth of IMBHs in the early Universe.
If successful, this would not only represent a breakthrough in our understanding of the epoch of reionization,
but it would also shed some light on the puzzles concerning the formation of SMBHs at $z\sim7$.

\section{Discussion and conclusions}

In this paper, basing on recent observational results,
we have studied the reionization history of cosmic hydrogen and helium
by both mini-quasars and stars. The contribution from quasars is also included.
In order to take into account the contribution from (mini-)quasars, the
$M_{\mathrm{bh}}-\sigma_{*}$ relation at $z>6$, which is from
the simulation on the merger tree of dark matter halos, is applied.
Several types of power-law spectra for the mini-quasar emission are considered.
Three different escape fractions $f_{\mathrm{esc}}$ of ionizing photons from stars are tested.
By simultaneously considering the constraints from the CMB
optical depth and observations of the Gunn-Peterson
trough in the spectra of high-redshift quasars, we find the following results for models with
a power-law IMBH spectrum (spectral index $\alpha=1$).
\begin{enumerate}
\item For mini-quasars with no UV ionizing photons ($200\,\mathrm{eV}<E<100\,\mathrm{keV}$),
the hydrogen ionizing photons emitted by mini-quasars are negligible.
Helium reionization occurs at $z\approx2.7$, consistent with observational results.
Only the model with $f_{\mathrm{esc}}\approx0.5$ is simultaneously consistent with
the constraints from the Gunn-Peterson test and the CMB optical depth.
\item When considering mini-quasars with UV ionizing photons ($10.4\,\mathrm{eV}<E<100\,\mathrm{keV}$),
both hydrogen and helium reionization are completed earlier.
In this case, an escape fraction in the range $0.2<f_{\mathrm{esc}}<0.5$ is able to satisfy the joint constraints.
Note that, in this optimistic case, mini-quasars are able to contribute $\sim20\%$ of hydrogen ionizing photons, consistent
with previous studies \citep{2004ApJ...604..484M,2010RAA....10..199W}.
\end{enumerate}
We also consider the possibility that the mini-quasar has a quasar-like spectral index of $\alpha=1.7$.
In this scenario, mini-quasars significantly over-produce
the helium ionizing photons, leading to an early completion of
helium reionization at $z\gtrsim4$, which is inconsistent with observations.

There are some caveats which could significantly affect our expectations about the contribution to cosmic reionization from mini-quasars.
First, we have assumed that all the ionizing photons from mini-quasars are
available for ionizing the IGM. This is because the high production
rate of ionizing photons from mini-quasars would allow the surrounding gas
to be ionized very quickly \citep[e.g.][]{2004ApJ...610...14W}. However, \citet{2013ApJ...770...76B}
argued that the escape fraction of ionizing photons
from the early quasars could be only $f_{\mathrm{esc,BH}}\sim0.3$.
If this is the case, the contribution from mini-quasars would be significantly
reduced. Another important consideration is the assumption on
the duty cycle of mini-quasars at high redshifts, i.\,e. $P(z>6)\sim0.9-1$. If the duty cycle
of mini-quasars at high redshifts is similar to that of present quasars ($P(z\sim0)\sim0.1)$,
the contribution of mini-quasars to cosmic reionization would become
negligible. Besides, on the assumption of the spectrum of mini-quasars,
instead of a simple power-law, one can also adopt a multicomponent spectrum
\citep[e.g.][]{2012MNRAS.425.2974T}.
Also note that we have ignored the secondary ionizations from the mini-quasar emission in our model for simplicity.
We caution that, until most of these properties of mini-quasars have been better understood,
a stringent constraint on the contribution of mini-quasars can not be obtained.
In turn, it is also reasonable to expect that,
improved measurements of helium reionization, especially at high redshifts,
coupled with the constraint from the SXRB, could be used to better constrain these properties of mini-quasars,
which would make a relevant contribution to the solution of the puzzles related to the formation of SMBHs at high redshifts.

Moreover, it is commonly believed that the feedback from the accretion onto
central BHs could play an important role.
For example, the radiation from accreting BHs can limit the gas supply by radiation pressures and photon-heating,
leading to the self-regulation of the BH growth \citep[e.g.][]{2012ApJ...754...34J,2012ApJ...747....9P}.
Additionally, \citet{2012MNRAS.425.2974T} suggest that the IGM heating by mini-quasars can also suppress
the formation and growth of central BHs in low-mass halos.
On the other hand, the star formation in low-mass galaxies may be affected by both local and global heating from mini-quasars
\citep[e.g.][]{2008MNRAS.387..158R,2012ApJ...754...34J}.
This would thus influence the relative contributions of mini-quasars and
stars to cosmic reionization. However, this additional complication
is beyond the scope of this paper, and will be addressed in our future work.

\section*{Acknowledgments}
We would like to thank the anonymous referee for her/his very helpful suggestions that
significantly improved the manuscript.
This work is partially supported by
National Basic Research Program of China (2009CB824800, 2012CB821800),
the National Natural Science Foundation (11073020, 11133005, 11233003),
and the Fundamental Research Funds for the Central Universities (WK2030220004).


\begin{thebibliography}{}

\bibitem[\protect\citeauthoryear{{Alvarez}, {Finlator} \& {Trenti}}{{Alvarez}
  et~al.}{2012}]{2012ApJ...759L..38A}
{Alvarez} M.~A.,  {Finlator} K.,    {Trenti} M.,  2012, \apjl, 759, L38

\bibitem[\protect\citeauthoryear{{Barkana} \& {Loeb}}{{Barkana} \&
  {Loeb}}{2001}]{2001PhR...349..125B}
{Barkana} R.,  {Loeb} A.,  2001, \physrep, 349, 125

\bibitem[\protect\citeauthoryear{{Becker}, {Bolton}, {Haehnelt} \&
  {Sargent}}{{Becker} et~al.}{2011}]{2011MNRAS.410.1096B}
{Becker} G.~D.,  {Bolton} J.~S.,  {Haehnelt} M.~G.,    {Sargent} W.~L.~W.,
  2011, \mnras, 410, 1096

\bibitem[\protect\citeauthoryear{{Benson}, {Venkatesan} \& {Shull}}{{Benson}
  et~al.}{2013}]{2013ApJ...770...76B}
{Benson} A.,  {Venkatesan} A.,    {Shull} J.~M.,  2013, \apj, 770, 76

\bibitem[\protect\citeauthoryear{{Bolton} \& {Haehnelt}}{{Bolton} \&
  {Haehnelt}}{2007}]{2007MNRAS.382..325B}
{Bolton} J.~S.,  {Haehnelt} M.~G.,  2007, \mnras, 382, 325

\bibitem[\protect\citeauthoryear{{Bolton}, {Haehnelt}, {Warren}, {Hewett},
  {Mortlock}, {Venemans}, {McMahon} \& {Simpson}}{{Bolton}
  et~al.}{2011}]{2011MNRAS.416L..70B}
{Bolton} J.~S.,  {Haehnelt} M.~G.,  {Warren} S.~J.,  {Hewett} P.~C.,
  {Mortlock} D.~J.,  {Venemans} B.~P.,  {McMahon} R.~G.,    {Simpson} C.,
  2011, \mnras, 416, L70

\bibitem[\protect\citeauthoryear{{Bouwens}, {Illingworth}, {Franx} \&
  {Ford}}{{Bouwens} et~al.}{2008}]{2008ApJ...686..230B}
{Bouwens} R.~J.,  {Illingworth} G.~D.,  {Franx} M.,    {Ford} H.,  2008, \apj,
  686, 230

\bibitem[\protect\citeauthoryear{{Bouwens} et~al.,}{{Bouwens}
  et~al.}{2012}]{2012ApJ...752L...5B}
{Bouwens} R.~J.  et~al., 2012, \apjl, 752, L5

\bibitem[\protect\citeauthoryear{{Bromm} \& {Loeb}}{{Bromm} \&
  {Loeb}}{2003}]{2003ApJ...596...34B}
{Bromm} V.,  {Loeb} A.,  2003, \apj, 596, 34

\bibitem[\protect\citeauthoryear{{Bunker}, {Stanway}, {Ellis} \&
  {McMahon}}{{Bunker} et~al.}{2004}]{2004MNRAS.355..374B}
{Bunker} A.~J.,  {Stanway} E.~R.,  {Ellis} R.~S.,    {McMahon} R.~G.,  2004,
  \mnras, 355, 374

\bibitem[\protect\citeauthoryear{{Bunker} et~al.,}{{Bunker}
  et~al.}{2010}]{2010MNRAS.409..855B}
{Bunker} A.~J.  et~al., 2010, \mnras, 409, 855

\bibitem[\protect\citeauthoryear{{Choudhury} \& {Ferrara}}{{Choudhury} \&
  {Ferrara}}{2005}]{2005MNRAS.361..577C}
{Choudhury} T.~R.,  {Ferrara} A.,  2005, \mnras, 361, 577

\bibitem[\protect\citeauthoryear{{Choudhury} \& {Ferrara}}{{Choudhury} \&
  {Ferrara}}{2006}]{2006MNRAS.371L..55C}
{Choudhury} T.~R.,  {Ferrara} A.,  2006, \mnras, 371, L55

\bibitem[\protect\citeauthoryear{{Cowie}, {Barger} \& {Trouille}}{{Cowie}
  et~al.}{2009}]{2009ApJ...692.1476C}
{Cowie} L.~L.,  {Barger} A.~J.,    {Trouille} L.,  2009, \apj, 692, 1476

\bibitem[\protect\citeauthoryear{{Dijkstra}}{{Dijkstra}}{2006}]{2006NewAR..50.%
.204D}
{Dijkstra} M.,  2006, \nar, 50, 204

\bibitem[\protect\citeauthoryear{{Dijkstra}, {Haiman} \& {Loeb}}{{Dijkstra}
  et~al.}{2004}]{2004ApJ...613..646D}
{Dijkstra} M.,  {Haiman} Z.,    {Loeb} A.,  2004, \apj, 613, 646

\bibitem[\protect\citeauthoryear{{Fan}, {Carilli} \& {Keating}}{{Fan}
  et~al.}{2006}]{2006ARA&A..44..415F}
{Fan} X.,  {Carilli} C.~L.,    {Keating} B.,  2006, \araa, 44, 415

\bibitem[\protect\citeauthoryear{{Fan} et~al.,}{{Fan}
  et~al.}{2001}]{2001AJ....122.2833F}
{Fan} X.  et~al., 2001, \aj, 122, 2833

\bibitem[\protect\citeauthoryear{{Farrell}, {Webb}, {Barret}, {Godet} \&
  {Rodrigues}}{{Farrell} et~al.}{2009}]{2009Natur.460...73F}
{Farrell} S.~A.,  {Webb} N.~A.,  {Barret} D.,  {Godet} O.,    {Rodrigues}
  J.~M.,  2009, \nat, 460, 73

\bibitem[\protect\citeauthoryear{{Faucher-Gigu{\`e}re}, {Lidz}, {Hernquist} \&
  {Zaldarriaga}}{{Faucher-Gigu{\`e}re} et~al.}{2008}]{2008ApJ...682L...9F}
{Faucher-Gigu{\`e}re} C.-A.,  {Lidz} A.,  {Hernquist} L.,    {Zaldarriaga} M.,
  2008, \apjl, 682, L9

\bibitem[\protect\citeauthoryear{{Ferrara} \& {Loeb}}{{Ferrara} \&
  {Loeb}}{2013}]{2013MNRAS.431.2826F}
{Ferrara} A.,  {Loeb} A.,  2013, \mnras, 431, 2826

\bibitem[\protect\citeauthoryear{{Ferrarese}}{{Ferrarese}}{2002}]{2002ApJ...57%
8...90F}
{Ferrarese} L.,  2002, \apj, 578, 90

\bibitem[\protect\citeauthoryear{{Finkelstein} et~al.,}{{Finkelstein}
  et~al.}{2012}]{2012ApJ...758...93F}
{Finkelstein} S.~L.  et~al., 2012, \apj, 758, 93

\bibitem[\protect\citeauthoryear{{Fontanot}, {Cristiani}, {Pfrommer}, {Cupani}
  \& {Vanzella}}{{Fontanot} et~al.}{2014}]{2014MNRAS.438.2097F}
{Fontanot} F.,  {Cristiani} S.,  {Pfrommer} C.,  {Cupani} G.,    {Vanzella} E.,
   2014, \mnras, 438, 2097

\bibitem[\protect\citeauthoryear{{Fontanot}, {Cristiani} \&
  {Vanzella}}{{Fontanot} et~al.}{2012}]{2012MNRAS.425.1413F}
{Fontanot} F.,  {Cristiani} S.,    {Vanzella} E.,  2012, \mnras, 425, 1413

\bibitem[\protect\citeauthoryear{{Furlanetto} \& {Dixon}}{{Furlanetto} \&
  {Dixon}}{2010}]{2010ApJ...714..355F}
{Furlanetto} S.~R.,  {Dixon} K.~L.,  2010, \apj, 714, 355

\bibitem[\protect\citeauthoryear{{Giavalisco} et~al.,}{{Giavalisco}
  et~al.}{2004}]{2004ApJ...600L.103G}
{Giavalisco} M.  et~al., 2004, \apjl, 600, L103

\bibitem[\protect\citeauthoryear{{Haardt} \& {Madau}}{{Haardt} \&
  {Madau}}{2012}]{2012ApJ...746..125H}
{Haardt} F.,  {Madau} P.,  2012, \apj, 746, 125

\bibitem[\protect\citeauthoryear{{Haiman} \& {Bryan}}{{Haiman} \&
  {Bryan}}{2006}]{2006ApJ...650....7H}
{Haiman} Z.,  {Bryan} G.~L.,  2006, \apj, 650, 7

\bibitem[\protect\citeauthoryear{{Haiman} \& {Loeb}}{{Haiman} \&
  {Loeb}}{2001}]{2001ApJ...552..459H}
{Haiman} Z.,  {Loeb} A.,  2001, \apj, 552, 459

\bibitem[\protect\citeauthoryear{{Heckman}, {Sembach}, {Meurer}, {Leitherer},
  {Calzetti} \& {Martin}}{{Heckman} et~al.}{2001}]{2001ApJ...558...56H}
{Heckman} T.~M.,  {Sembach} K.~R.,  {Meurer} G.~R.,  {Leitherer} C.,
  {Calzetti} D.,    {Martin} C.~L.,  2001, \apj, 558, 56

\bibitem[\protect\citeauthoryear{{Hopkins} \& {Beacom}}{{Hopkins} \&
  {Beacom}}{2006}]{2006ApJ...651..142H}
{Hopkins} A.~M.,  {Beacom} J.~F.,  2006, \apj, 651, 142

\bibitem[\protect\citeauthoryear{{Iliev}, {Mellema}, {Shapiro} \&
  {Pen}}{{Iliev} et~al.}{2007}]{2007MNRAS.376..534I}
{Iliev} I.~T.,  {Mellema} G.,  {Shapiro} P.~R.,    {Pen} U.-L.,  2007, \mnras,
  376, 534

\bibitem[\protect\citeauthoryear{{Iwata} et~al.,}{{Iwata}
  et~al.}{2009}]{2009ApJ...692.1287I}
{Iwata} I.  et~al., 2009, \apj, 692, 1287

\bibitem[\protect\citeauthoryear{{Jeon}, {Pawlik}, {Greif}, {Glover}, {Bromm},
  {Milosavljevi{\'c}} \& {Klessen}}{{Jeon} et~al.}{2012}]{2012ApJ...754...34J}
{Jeon} M.,  {Pawlik} A.~H.,  {Greif} T.~H.,  {Glover} S.~C.~O.,  {Bromm} V.,
  {Milosavljevi{\'c}} M.,    {Klessen} R.~S.,  2012, \apj, 754, 34

\bibitem[\protect\citeauthoryear{{Komatsu} et~al.,}{{Komatsu}
  et~al.}{2011}]{2011ApJS..192...18K}
{Komatsu} E.  et~al., 2011, \apjs, 192, 18

\bibitem[\protect\citeauthoryear{{Krumholz} \& {Dekel}}{{Krumholz} \&
  {Dekel}}{2012}]{2012ApJ...753...16K}
{Krumholz} M.~R.,  {Dekel} A.,  2012, \apj, 753, 16

\bibitem[\protect\citeauthoryear{{Kuhlen} \& {Faucher-Gigu{\`e}re}}{{Kuhlen} \&
  {Faucher-Gigu{\`e}re}}{2012}]{2012MNRAS.423..862K}
{Kuhlen} M.,  {Faucher-Gigu{\`e}re} C.-A.,  2012, \mnras, 423, 862

\bibitem[\protect\citeauthoryear{{Kuhlen}, {Krumholz}, {Madau}, {Smith} \&
  {Wise}}{{Kuhlen} et~al.}{2012}]{2012ApJ...749...36K}
{Kuhlen} M.,  {Krumholz} M.~R.,  {Madau} P.,  {Smith} B.~D.,    {Wise} J.,
  2012, \apj, 749, 36

\bibitem[\protect\citeauthoryear{{Kuhlen} \& {Madau}}{{Kuhlen} \&
  {Madau}}{2005}]{2005MNRAS.363.1069K}
{Kuhlen} M.,  {Madau} P.,  2005, \mnras, 363, 1069

\bibitem[\protect\citeauthoryear{{Li}}{{Li}}{2008}]{2008MNRAS.388.1487L}
{Li} L.-X.,  2008, \mnras, 388, 1487

\bibitem[\protect\citeauthoryear{{Lodato} \& {Natarajan}}{{Lodato} \&
  {Natarajan}}{2006}]{2006MNRAS.371.1813L}
{Lodato} G.,  {Natarajan} P.,  2006, \mnras, 371, 1813

\bibitem[\protect\citeauthoryear{{L{\"u}tzgendorf} et~al.,}{{L{\"u}tzgendorf}
  et~al.}{2013}]{2013A&A...552A..49L}
{L{\"u}tzgendorf} N.  et~al., 2013, \aap, 552, A49

\bibitem[\protect\citeauthoryear{{Madau} \& {Rees}}{{Madau} \&
  {Rees}}{2001}]{2001ApJ...551L..27M}
{Madau} P.,  {Rees} M.~J.,  2001, \apjl, 551, L27

\bibitem[\protect\citeauthoryear{{Madau}, {Rees}, {Volonteri}, {Haardt} \&
  {Oh}}{{Madau} et~al.}{2004}]{2004ApJ...604..484M}
{Madau} P.,  {Rees} M.~J.,  {Volonteri} M.,  {Haardt} F.,    {Oh} S.~P.,  2004,
  \apj, 604, 484

\bibitem[\protect\citeauthoryear{{McQuinn}}{{McQuinn}}{2009}]{2009ApJ...704L..%
89M}
{McQuinn} M.,  2009, \apjl, 704, L89

\bibitem[\protect\citeauthoryear{{McQuinn}}{{McQuinn}}{2012}]{2012MNRAS.426.13%
49M}
{McQuinn} M.,  2012, \mnras, 426, 1349

\bibitem[\protect\citeauthoryear{{Meiksin} \& {Tittley}}{{Meiksin} \&
  {Tittley}}{2012}]{2012MNRAS.423....7M}
{Meiksin} A.,  {Tittley} E.~R.,  2012, \mnras, 423, 7

\bibitem[\protect\citeauthoryear{{Miller}, {Fabbiano}, {Miller} \&
  {Fabian}}{{Miller} et~al.}{2003}]{2003ApJ...585L..37M}
{Miller} J.~M.,  {Fabbiano} G.,  {Miller} M.~C.,    {Fabian} A.~C.,  2003,
  \apjl, 585, L37

\bibitem[\protect\citeauthoryear{{Mitra}, {Ferrara} \& {Choudhury}}{{Mitra}
  et~al.}{2013}]{2013MNRAS.428L...1M}
{Mitra} S.,  {Ferrara} A.,    {Choudhury} T.~R.,  2013, \mnras, 428, L1

\bibitem[\protect\citeauthoryear{{Mitsuda} et~al.,}{{Mitsuda}
  et~al.}{1984}]{1984PASJ...36..741M}
{Mitsuda} K.  et~al., 1984, \pasj, 36, 741

\bibitem[\protect\citeauthoryear{{Mortlock} et~al.,}{{Mortlock}
  et~al.}{2011}]{2011Natur.474..616M}
{Mortlock} D.~J.  et~al., 2011, \nat, 474, 616

\bibitem[\protect\citeauthoryear{{Nestor}, {Shapley}, {Steidel} \&
  {Siana}}{{Nestor} et~al.}{2011}]{2011ApJ...736...18N}
{Nestor} D.~B.,  {Shapley} A.~E.,  {Steidel} C.~C.,    {Siana} B.,  2011, \apj,
  736, 18

\bibitem[\protect\citeauthoryear{{Oh} \& {Haiman}}{{Oh} \&
  {Haiman}}{2002}]{2002ApJ...569..558O}
{Oh} S.~P.,  {Haiman} Z.,  2002, \apj, 569, 558

\bibitem[\protect\citeauthoryear{{Ouchi} et~al.,}{{Ouchi}
  et~al.}{2010}]{2010ApJ...723..869O}
{Ouchi} M.  et~al., 2010, \apj, 723, 869

\bibitem[\protect\citeauthoryear{{Ouchi} et~al.,}{{Ouchi}
  et~al.}{2004}]{2004ApJ...611..660O}
{Ouchi} M.  et~al., 2004, \apj, 611, 660

\bibitem[\protect\citeauthoryear{{Paardekooper}, {Pelupessy}, {Altay} \&
  {Kruip}}{{Paardekooper} et~al.}{2011}]{2011A&A...530A..87P}
{Paardekooper} J.-P.,  {Pelupessy} F.~I.,  {Altay} G.,    {Kruip} C.~J.~H.,
  2011, \aap, 530, A87

\bibitem[\protect\citeauthoryear{{Park} \& {Ricotti}}{{Park} \&
  {Ricotti}}{2012}]{2012ApJ...747....9P}
{Park} K.,  {Ricotti} M.,  2012, \apj, 747, 9

\bibitem[\protect\citeauthoryear{{Pizzella}, {Corsini}, {Dalla Bont{\`a}},
  {Sarzi}, {Coccato} \& {Bertola}}{{Pizzella}
  et~al.}{2005}]{2005ApJ...631..785P}
{Pizzella} A.,  {Corsini} E.~M.,  {Dalla Bont{\`a}} E.,  {Sarzi} M.,  {Coccato}
  L.,    {Bertola} F.,  2005, \apj, 631, 785

\bibitem[\protect\citeauthoryear{{Razoumov} \& {Sommer-Larsen}}{{Razoumov} \&
  {Sommer-Larsen}}{2006}]{2006ApJ...651L..89R}
{Razoumov} A.~O.,  {Sommer-Larsen} J.,  2006, \apjl, 651, L89

\bibitem[\protect\citeauthoryear{{Reddy}, {Steidel}, {Pettini}, {Adelberger},
  {Shapley}, {Erb} \& {Dickinson}}{{Reddy} et~al.}{2008}]{2008ApJS..175...48R}
{Reddy} N.~A.,  {Steidel} C.~C.,  {Pettini} M.,  {Adelberger} K.~L.,  {Shapley}
  A.~E.,  {Erb} D.~K.,    {Dickinson} M.,  2008, \apjs, 175, 48

\bibitem[\protect\citeauthoryear{{Ripamonti}, {Mapelli} \&
  {Zaroubi}}{{Ripamonti} et~al.}{2008}]{2008MNRAS.387..158R}
{Ripamonti} E.,  {Mapelli} M.,    {Zaroubi} S.,  2008, \mnras, 387, 158

\bibitem[\protect\citeauthoryear{{Robertson}, {Ellis}, {Dunlop}, {McLure} \&
  {Stark}}{{Robertson} et~al.}{2010}]{2010Natur.468...49R}
{Robertson} B.~E.,  {Ellis} R.~S.,  {Dunlop} J.~S.,  {McLure} R.~J.,    {Stark}
  D.~P.,  2010, \nat, 468, 49

\bibitem[\protect\citeauthoryear{{Robertson} et~al.,}{{Robertson}
  et~al.}{2013}]{2013ApJ...768...71R}
{Robertson} B.~E.  et~al., 2013, \apj, 768, 71

\bibitem[\protect\citeauthoryear{{Salvaterra}, {Haardt} \&
  {Ferrara}}{{Salvaterra} et~al.}{2005}]{2005MNRAS.362L..50S}
{Salvaterra} R.,  {Haardt} F.,    {Ferrara} A.,  2005, \mnras, 362, L50

\bibitem[\protect\citeauthoryear{{Schenker}, {Stark}, {Ellis}, {Robertson},
  {Dunlop}, {McLure}, {Kneib} \& {Richard}}{{Schenker}
  et~al.}{2012}]{2012ApJ...744..179S}
{Schenker} M.~A.,  {Stark} D.~P.,  {Ellis} R.~S.,  {Robertson} B.~E.,  {Dunlop}
  J.~S.,  {McLure} R.~J.,  {Kneib} J.-P.,    {Richard} J.,  2012, \apj, 744,
  179

\bibitem[\protect\citeauthoryear{{Schroeder}, {Mesinger} \&
  {Haiman}}{{Schroeder} et~al.}{2013}]{2013MNRAS.428.3058S}
{Schroeder} J.,  {Mesinger} A.,    {Haiman} Z.,  2013, \mnras, 428, 3058

\bibitem[\protect\citeauthoryear{{Shankar}, {Crocce}, {Miralda-Escud{\'e}},
  {Fosalba} \& {Weinberg}}{{Shankar} et~al.}{2010}]{2010ApJ...718..231S}
{Shankar} F.,  {Crocce} M.,  {Miralda-Escud{\'e}} J.,  {Fosalba} P.,
  {Weinberg} D.~H.,  2010, \apj, 718, 231

\bibitem[\protect\citeauthoryear{{Shen} et~al.,}{{Shen}
  et~al.}{2007}]{2007AJ....133.2222S}
{Shen} Y.  et~al., 2007, \aj, 133, 2222

\bibitem[\protect\citeauthoryear{{Sheth} \& {Tormen}}{{Sheth} \&
  {Tormen}}{1999}]{1999MNRAS.308..119S}
{Sheth} R.~K.,  {Tormen} G.,  1999, \mnras, 308, 119

\bibitem[\protect\citeauthoryear{{Shull}, {France}, {Danforth}, {Smith} \&
  {Tumlinson}}{{Shull} et~al.}{2010}]{2010ApJ...722.1312S}
{Shull} J.~M.,  {France} K.,  {Danforth} C.~W.,  {Smith} B.,    {Tumlinson} J.,
   2010, \apj, 722, 1312

\bibitem[\protect\citeauthoryear{{Tanaka} \& {Haiman}}{{Tanaka} \&
  {Haiman}}{2009}]{2009ApJ...696.1798T}
{Tanaka} T.,  {Haiman} Z.,  2009, \apj, 696, 1798

\bibitem[\protect\citeauthoryear{{Tanaka}, {Perna} \& {Haiman}}{{Tanaka}
  et~al.}{2012}]{2012MNRAS.425.2974T}
{Tanaka} T.,  {Perna} R.,    {Haiman} Z.,  2012, \mnras, 425, 2974

\bibitem[\protect\citeauthoryear{{Tanaka}}{{Tanaka}}{2014}]{2014CQGra..31x4005%
T}
{Tanaka} T.~L.,  2014, Classical and Quantum Gravity, 31, 244005

\bibitem[\protect\citeauthoryear{{Tanaka} \& {Li}}{{Tanaka} \&
  {Li}}{2014}]{2014MNRAS.439.1092T}
{Tanaka} T.~L.,  {Li} M.,  2014, \mnras, 439, 1092

\bibitem[\protect\citeauthoryear{{Telfer}, {Zheng}, {Kriss} \&
  {Davidsen}}{{Telfer} et~al.}{2002}]{2002ApJ...565..773T}
{Telfer} R.~C.,  {Zheng} W.,  {Kriss} G.~A.,    {Davidsen} A.~F.,  2002, \apj,
  565, 773

\bibitem[\protect\citeauthoryear{{Tozzi} et~al.,}{{Tozzi}
  et~al.}{2006}]{2006A&A...451..457T}
{Tozzi} P.  et~al., 2006, \aap, 451, 457

\bibitem[\protect\citeauthoryear{{Trenti}, {Stiavelli}, {Bouwens}, {Oesch},
  {Shull}, {Illingworth}, {Bradley} \& {Carollo}}{{Trenti}
  et~al.}{2010}]{2010ApJ...714L.202T}
{Trenti} M.,  {Stiavelli} M.,  {Bouwens} R.~J.,  {Oesch} P.,  {Shull} J.~M.,
  {Illingworth} G.~D.,  {Bradley} L.~D.,    {Carollo} C.~M.,  2010, \apjl, 714,
  L202

\bibitem[\protect\citeauthoryear{{Vanzella} et~al.,}{{Vanzella}
  et~al.}{2010}]{2010ApJ...725.1011V}
{Vanzella} E.  et~al., 2010, \apj, 725, 1011

\bibitem[\protect\citeauthoryear{{Vanzella} et~al.,}{{Vanzella}
  et~al.}{2012}]{2012ApJ...751...70V}
{Vanzella} E.  et~al., 2012, \apj, 751, 70

\bibitem[\protect\citeauthoryear{{Volonteri}, {Haardt} \& {Madau}}{{Volonteri}
  et~al.}{2003}]{2003ApJ...582..559V}
{Volonteri} M.,  {Haardt} F.,    {Madau} P.,  2003, \apj, 582, 559

\bibitem[\protect\citeauthoryear{{Wang}, {Chen} \& {Zhang}}{{Wang}
  et~al.}{2006}]{2006ApJ...647L..17W}
{Wang} J.-M.,  {Chen} Y.-M.,    {Zhang} F.,  2006, \apjl, 647, L17

\bibitem[\protect\citeauthoryear{{Wang}, {Wang}, {Xiang}, {Wang}, {Hao} \&
  {Yuan}}{{Wang} et~al.}{2010}]{2010RAA....10..199W}
{Wang} Y.-Y.,  {Wang} L.,  {Xiang} S.-P.,  {Wang} Y.,  {Hao} J.-M.,    {Yuan}
  Y.-F.,  2010, Research in Astronomy and Astrophysics, 10, 199

\bibitem[\protect\citeauthoryear{{Whalen}, {Abel} \& {Norman}}{{Whalen}
  et~al.}{2004}]{2004ApJ...610...14W}
{Whalen} D.,  {Abel} T.,    {Norman} M.~L.,  2004, \apj, 610, 14

\bibitem[\protect\citeauthoryear{{Willott} et~al.,}{{Willott}
  et~al.}{2010}]{2010AJ....139..906W}
{Willott} C.~J.  et~al., 2010, \aj, 139, 906

\bibitem[\protect\citeauthoryear{{Wise}, {Demchenko}, {Halicek}, {Norman},
  {Turk}, {Abel} \& {Smith}}{{Wise} et~al.}{2014}]{2014MNRAS.442.2560W}
{Wise} J.~H.,  {Demchenko} V.~G.,  {Halicek} M.~T.,  {Norman} M.~L.,  {Turk}
  M.~J.,  {Abel} T.,    {Smith} B.~D.,  2014, \mnras, 442, 2560

\bibitem[\protect\citeauthoryear{{Worseck} et~al.,}{{Worseck}
  et~al.}{2011}]{2011ApJ...733L..24W}
{Worseck} G.  et~al., 2011, \apjl, 733, L24

\bibitem[\protect\citeauthoryear{{Wyithe} \& {Loeb}}{{Wyithe} \&
  {Loeb}}{2003}]{2003ApJ...586..693W}
{Wyithe} J.~S.~B.,  {Loeb} A.,  2003, \apj, 586, 693

\bibitem[\protect\citeauthoryear{{Yuan}, {Zhang}, {Yang}, {Wang}, {Lin}, {Xu}
  \& {Zhang}}{{Yuan} et~al.}{2008}]{2008AIPC..968..369Y}
{Yuan} Y.-F.,  {Zhang} C.-Y.,  {Yang} J.-M.,  {Wang} L.,  {Lin} X.-B.,  {Xu}
  S.-N.,    {Zhang} J.-L.,  2008, in {Yuan} Y.-F.,  {Li} X.-D.,   {Lai} D.,
  eds,  American Institute of Physics Conference Series Vol. 968, Astrophysics
  of Compact Objects. pp 369--374

\bibitem[\protect\citeauthoryear{{Zaroubi}, {Thomas}, {Sugiyama} \&
  {Silk}}{{Zaroubi} et~al.}{2007}]{2007MNRAS.375.1269Z}
{Zaroubi} S.,  {Thomas} R.~M.,  {Sugiyama} N.,    {Silk} J.,  2007, \mnras,
  375, 1269

\bibitem[\protect\citeauthoryear{{Zheng} et~al.,}{{Zheng}
  et~al.}{2004}]{2004ApJ...605..631Z}
{Zheng} W.  et~al., 2004, \apj, 605, 631

\end{thebibliography}

\clearpage{}

\label{lastpage}

\end{document}